# What drives the acceptance of AI technology?

## : the role of expectations and experiences

First Author Minsang Yi

Ph.D., Seoul National University, Republic of Korea

Email: neoyms0@snu.ac.kr

Second Author Hanbyul Choi

Ph.D. Candidate, Seoul National University, Republic of Korea

Email: chb@snu.ac.kr

Abstract

In recent years, Artificial intelligence products and services have been offered potential users as pilots. The acceptance intention towards artificial intelligence is greatly influenced by the experience with current AI products and services, expectations for AI, and past experiences with ICT technology. This study aims to explore the factors that impact AI acceptance intention and understand the process of its formation. The analysis results of this study reveal that AI experience and past ICT experience affect AI acceptance intention in two ways. Through the direct path, higher AI experience and ICT experience are associated with a greater intention to accept AI. Additionally, there is an indirect path where AI experience and ICT experience contribute to increased expectations for AI, and these expectations, in turn, elevate acceptance intention. Based on the findings, several recommendations are suggested for companies and public organizations planning to implement artificial intelligence in the future. It is crucial to manage the user experience of ICT services and pilot AI products and services to deliver positive experiences. It is essential to provide potential AI users with specific information about the features and benefits of AI products and services. This will enable them to develop realistic expectations regarding AI technology.

I. introduction

As more people use information and communication technology, network effects come into play. For instance, social network platforms become more valuable as their user base grows. The greater the number of users, the more beneficial the platform becomes in terms of information, content, connectivity, and communication. These advantages attract new participants. Davis (1989) suggests that users' acceptance of technology is crucial for its successful adoption. Maruping et al. (2017) argue that the benefits of information and communication technology are realized through actual usage. Inappropriate use of ICT (such as not using it at all or using it inefficiently) remains a significant problem in practice. Venkatesh & Goyal (2010) identify the failure to ensure continuous usage of ICT technology as one of the causes of the productivity paradox. As more people embrace artificial intelligence, it will also generate greater personal and social benefits. Therefore, it is essential to examine the intention to accept artificial intelligence and the factors that influence it.

In recent years, AI services and products have been provided on a trial basis. The intention to adopt AI is highly likely to be influenced by the user's experience with currently available AI products and services, expectations of AI, and previous experience with ICT technologies. Through their current experiences with AI products and services, users can anticipate the potential changes their work and daily lives. And users have an opportunity to assess whether this technology can be beneficial to their lives. Also, users can form their intention to adopt next-generation technologies based on their previous successful utilization of ICT. If they have successfully utilized ICT in the past, they can expect to effectively utilize AI products and services as well, leading to a higher intention to adopt. Expectations regarding AI are formed based on experience, social context, and the image of the products and services.

This study discusses the intention to adopt AI and the factors that influence it. Specifically, it analyzes how trial experiences with AI products and services, previous experience with ICT, and expectations of AI will impact future intention to accept AI. Through this analysis, we can understand the process by how users' intention to accept AI is formed and discover important factors that influence this process. Based on the research results, we can propose approaches to manage user experience and expectations for the successful adoption of AI in various organizations.

This study is conducted as follows. First, it reviews theories and previous studies on technology acceptance, experience with technology, and expectations of technology. Second, it presents the models and research hypotheses for analyzing the relationships between technology acceptance, experience, and expectations. Third, it outlines the data and research method used in this study. Fourth, it presents descriptive statistics on key variables and the results of hypothesis testing. Finally, it provides a detailed discussion of the research findings and suggests recommendations.

II. Theoretical Background

1. Intention to Accept Artificial Intelligence Technology

Intention to accept refers to an individual's decision-making of accepting a specific technology. As AI develops and AI technology is provided to users, studies on the intention to introduce AI is being conducted in various fields. According to Lu, Cai & Gursoy (2019), research on the intention to accept AI has been conducted in areas such as fitness and health, smart homes, transportation, banking, smartphones, and e-commerce. Kelly et al. (2023) conducted a systematic literature review analyzing 60 studies on the adoption of AI technology. Previous studies have focused on customer service (27%), education (11%), healthcare (17%), organizations (15%), consumer products (15%), and other areas (15%). These studies demonstrate that AI implementation is progressing across multiple domains and that the acceptance of AI is recognized as a crucial common theme.

TAM (Technology Acceptance Model), EDT (Expectancy Disconfirmation Theory), and UTAUT (Unified Theory of Acceptance and Use of Technology) are theoretical frameworks that examine the factors influencing the intention to adopt technology, such as perceived usefulness, perceived ease of use, and expectancy-disconfirmation. According to TAM, external factors such as experience, media, and social support shape perceived usefulness (PU) and perceived ease of use (PEOU), which in turn influence the intention to accept technology and actual usage. In EDT, users form expectations about the technology before using it. After using the technology, they compare their actual experience with their prior expectations, and the degree of expectation fulfillment affects their continuous intention to use or adopt the technology. UTAUT incorporates TAM, EDT, and other eight theoretical acceptance models. According to this model, performance expectancy, social influence, effort expectancy, and facilitating conditions all influence the intention to accept and use technology. Sex, age, voluntariness of use, and prior experience moderate the effects of these factors on intention and behavior.

Studies on the intention to adopt AI-incorporated products and services also draw upon TAM, EDT, UTAUT, and other related theories (Gursoy et al., 2019). However, some criticism has been raised regarding previous studies and theories on the intention to adopt AI.

First, there is criticism that the intention to accept technology does not guarantee actual technology usage. An individual may have the intention to use a technology, but various obstacles or constraints may prevent them from actually using it. Maruping et al. (2017) argue that behavioral expectation should be considered alongside behavioral intention as a determining factor for technology acceptance. Behavioral expectation refers to an individual's subjective self-report on the probability of performing a specific behavior, based on their judgment of the factors influencing behavioral decisions.

Second, there is criticism that existing theories are not appropriate for analyzing the intention to accept artificial intelligence. The main point of this criticism is that existing models do not adequately reflect the characteristics of AI technology. Gursoy et al. (2019) criticize that existing models were developed for self-service technologies and do not consider the interaction between humans and AI. Lu, Cai & Gursoy (2019) criticize existing models for failing to consider the selection process between AI and human services.

Third, there is a point that the influence of experience on the intention to accept is not adequately considered. According to Lee (2011), prior experience with a technology affects technology acceptance and continued usage by providing an opportunity to evaluate satisfaction with the technology. It is also necessary to consider how prior ICT experiences influence the intention to accept AI. Martins & Kambil (1999) found that previous successful utilization of ICT has a positive impact on the adoption of subsequent ICT. In the context of AI, it can be expected that users who are familiar with and actively utilize ICT will also be proactive in adopting AI.

Fourth, there is a limitation in that the expectations of technology are limited to specific contexts such in work and task performance. TAM presents perceived usefulness and perceived ease of use as factors of expectations for technology, while UTAUT presents performance expectations and effort expectations. However, currently, AI is being provided on a trial basis, and the exploration of its application methods and domains is ongoing. In this context, current AI expectations may be about AI-driven changes in life, changes in society, etc., rather than specific expectations for work and tasks. Why Common Expectations, Images, and Beliefs About AI Matter Au-Yong-Oliveira et al. (2020) suggests. The study presents expectations of AI as the public's hopes for this technology. Expectations of AI reduce aversion to AI, algorithms, robots, etc., and determine generosity towards AI services.

2. Experience with Artificial Intelligence

Experience refers to knowledge, feelings, memories, evaluations, skills, etc. acquired through participation in or observation of events, activities, and situations. Experience with technology can be defined as the knowledge, emotions, memories, evaluations, and so on acquired through direct usage and exposure to a technology. Experience can influence cognition and behavior. It is also expected to have an impact on the intention to accept a technology and expectations towards it. Taylor & Todd (1995) suggest that past experience is a significant determinant of behavioral intention. Knowledge gained from past experiences is more easily remembered, and the events experienced receive greater emphasis in the intention formation process. In the process of forming the intention to accept ICT, the level of ICT experience acts as a moderating variable in influencing individual factors.

Experience related to artificial intelligence can be divided into two categories. Firstly, there is experience with AI products and services. Knowledge, appraisal, and evaluation of AI can be acquired through the use of AI products and services. These will influence AI adoption intentions and expectations. In a survey conducted by Pegasystems Inc (2017) among the public, only 25% of respondents who had never used AI answered that they would trust AI-operated companies. However, among respondents who had prior experience with AI, 55% provided positive responses. This indicates that AI usage experience has a positive influence on the perception of AI service providers and is also expected to have a positive impact on expectations towards AI. Studies analyzing the intention to accept autonomous vehicles have shown that the higher the familiarity with autonomous driving technology, the greater the exposure to related news articles, and the higher the level of prior knowledge, the stronger the intention to adopt autonomous vehicles becomes. Furthermore, concerns decrease, and there is a greater willingness to delegate driving control (Nees, 2016; Dai et al., 2021; Charness et al., 2018).

Research on previous ICT experiences is primarily found in the field of education. According to Osborn (2001), students who have previous computer experience and are confident in their computing skills are more likely to experience less anxiety and achieve better results when using computers for learning. Holden & Weeden (2003) compared students with and without previous programming experience in programming-related courses. Students with experience achieved higher grades in certain subjects and had a higher completion rate for the course. Shih et al. (2006) showed that participants with more experience in internet utilized time more efficiently and behaved more effective in web-based educational courses. Lin (2011) presented results that users with previous experience had a higher intention to use e-learning systems. On the other hand, Leong et al. (2018) found that students who were more familiar with mobile social networking services (SNS) had more difficulty in being motivated to use educational applications for learning activities, which they attributed to the experiences of using mobile SNS for pleasure and entertainment purposes.

3. Expectations for Artificial Intelligence

Expectations can be defined as a set of beliefs prior to experiencing a product or service. Expectations for information systems are related to users' beliefs regarding performance and system usage (Szajna & Scamell, 1993). The Expectation Disconfirmation Theory (EDT) explains how expectations influence usage intentions. When consumers are exposed to information about the performance characteristics of a product or service, specific beliefs or expectations are formed. Consumers cognitively compare their expectation levels with their actual post-usage experiences. The extent of the mismatch between expectations and experiences determines the level of product or service satisfaction, which in turn influences the intention for continued usage (Venkatesh & Goyal, 2010).

TAM (Technology Acceptance Model) and UTAUT (Unified Theory of Acceptance and Use of Technology) propose that perceived usefulness, perceived ease of use, performance expectations, and effort expectations are key determinants of technology acceptance. These factors represent the perceptions formed by users who have no prior experience with the technology. The measurement items for these factors are constructed similarly. Studies of these models generally suggest that higher expectations of performance (perceived usefulness, performance expectations) and easier usability (perceived ease of use, effort expectations) result in higher intentions to adopt the technology.

Expectations for technology not only affect individuals but also play a role in the development and adoption process of technology in society. According to Borup et al. (2006), expectations for technology are crucial as they serve as promises and visions of the technology, mobilizing macro and micro resources. Expectations for technology stimulate the interest of various actors, define their roles and establishing mutual obligations in the process of technology development and change. Szajna & Scamell (1993) express concerns that expectations for technology can be formed based on the technology's image and positive perception, unrelated to actual performance. It can be formed at unrealistic levels. Failing to meet unrealistic expectations can be a cause of information system failure of acceptance. To prevent the formation of unrealistic expectations, it is necessary to facilitate communication between users and developers, promote user participation in the development process, and make realistic promises that do not exceed what can be delivered.

Expectations about technology are closely related to experience. Oliver(1980) identified previous experience, brand, social context, seller persuasion, and individual characteristics as factors that can influence expectations. According to Taylor & Todd(1995), there are differences in expectation levels between users with experience and those without experience. Therefore, it is necessary to enable users without experience to form realistic expectations about technology. Bhattacherjee(2001) found that when users' expectations of online banking services are met by perceived performance after use, it improves satisfaction and the intention to continue using the information system. Additionally, it was found that user experience enhances perceived usefulness of the system. Initially, users may have a low level of perceived usefulness because they do not know what to expect from the information system. Through subsequent experiences, expectation levels are adjusted. This research proposes that to promote the continuous use of information systems, potential benefits of IS usage should be communicated to new users and effective ways of using IS should be taught to existing users to enhance their experience and satisfaction.

When the AI products are not met through experience, a decrease in product purchases is also observed. The Korea Consumer Agency conducted a survey in 2017 on the expectations and user experiences of AI speakers. The expected features before users purchased the product were identified as easy and convenient voice recognition (46.3%) and everyday conversation (23.0%). The discomfort with AI speakers was mainly related to inadequate voice recognition (56.7%), difficulties in connected conversations (45.7%), and misinterpretation of external noise as voice commands (37.0%). There was a gap between the level of expectations and the actual experience of AI speakers. After that Consumer Insight (2021) identified low satisfaction with AI speakers, including low accuracy of voice command recognition (47%), inability to engage in natural conversations (33%), and misinterpretation of external noise as voice commands (32.0%). These results were similar to the findings presented by the Korea Consumer Agency (2017). The mismatch between expectations and experiences regarding AI speakers remains unresolved. This disconfirmation between expectations and experiences has resulted in a decrease in frequency of use, satisfaction, and purchasing activities. The proportion of users who use AI speakers more than three times a week decreased from 53% in 2019 to 50% in 2021, and overall satisfaction decreased from 47% in 2019 to 41% in 2021. The proportion of direct purchases of AI speakers decreased from 69% in 2019 to 60% in 2021(Consumer Insight, 2021).

III. Research Model and Data

This study examines the impact of experiences and expectations regarding artificial intelligence on the intention to accept artificial intelligence. The definitions of each factor are as follows:

Artificial intelligence acceptance intention refers to the decision-making process in which an individual intends to purchase, use, or adopt artificial intelligence in the future. The experience of artificial intelligence products and services entails the evaluation that an individual presents after using artificial intelligence products and services. ICT utilization experience can be defined as a comprehensive cognitive judgment of the experiences (economy, leisure, information collection, communication, work, etc.) obtained through the use of ICT products and services (Limayem et al., 2007). Expectations of artificial intelligence pertain to the beliefs and convictions held by users of artificial intelligence products and services regarding the technology. Based on previous research and theory, the following hypotheses are posited regarding the relationship between AI experience, ICT experience, AI expectation, and acceptance intention.

H1. The more positive the AI experience, the higher the intention to accept AI.

H2. The more positive the evaluation of ICT utilization experience, the higher the intention to accept artificial intelligence.

H3. The higher the expectations of artificial intelligence, the higher the intention to accept artificial intelligence.

H4. Expectations of artificial intelligence will partially mediate the relationship between AI experience evaluation, ICT experience evaluation, and artificial intelligence acceptance intention.

[Figure 1] Research model

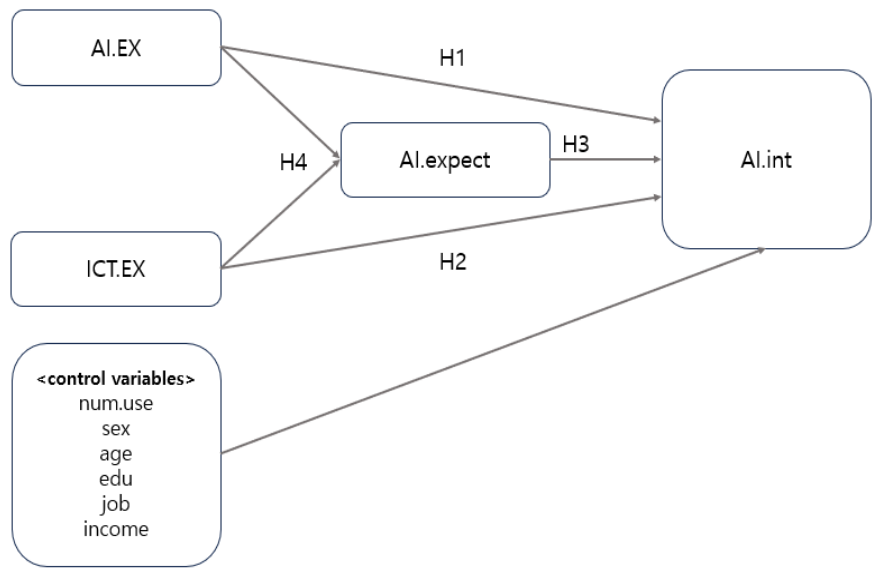

The research hypotheses were visually represented and presented in [Figure 1]. In this figure, AI.EX represents AI experience evaluation, ICT.EX represents ICT experience evaluation, AI.expect represents artificial intelligence expectations, and AI.int represents artificial intelligence acceptance intention. To accurately ascertain the relationships between AI experience, ICT experience, AI expectation, and acceptance intention, it is necessary to incorporate important demographic factors related to the digital divide into the model. The model includes control variables such as the number of AI products and services used (num.AI), gender (sex), age (age), highest level of education (edu), occupation (job), and average monthly household income (income) for each individual observation.

This study utilized data from the 2019 Digital Information Divide Fact-Finding Survey conducted by the National Information Society Agency in September 2019. This survey has been conducted annually since 2002, and the questionnaire items verified in the technology acceptance model were included in the survey. A layered probability proportional sampling method based on region was employed for sampling. This method selects survey participants while considering the proportion of each region in the total population. The dataset provided in this survey consists of a total of 7,000 individuals (National Information Society Agency, 2019). Among the 7,000 data points, this study focused on individuals who had prior experience using artificial intelligence products and services, and observations with missing information were excluded from the analysis. Consequently, the study analyzed the responses of a total of 2,622 participants.

<Table 1> Questionnaires by factor

| variables | no. | content |
|---|---|---|
| Intention to use AI AI.int | Q29A1 | Even if I have to pay more, I am willing to purchase a device or service if it incorporates artificial intelligence functions. |
| | Q29A2 | I strive to quickly adapt to artificial intelligence technology. |
| | Q29A3 | If a service with artificial intelligence technology is available, I actively use it. |
| | Q29A4 | I proactively recommend artificial intelligence technology devices or services to others. |
| AI experience AI.EX | Q24AC1 | A smartphone equipped with artificial intelligence is beneficial in my life. |
| | Q24AC2 | Artificial intelligence speakers are helpful in my life. |
| | Q25AC3 | Artificial intelligence voice assistants are helpful in my life. |
| ICT experience ICT.EX | Q20A1 | I have experienced increased happiness as I had more opportunities for leisure activities (hobbies, culture, and entertainment) through the use of digital devices. |
| | Q20A2 | Using digital devices, I am able to access and stay updated with news and information more quickly. |
| | Q20A3 | By utilizing digital devices, I have expanded my knowledge and acquired more information. |
| | Q20A4 | Digital devices have provided me with numerous opportunities to engage with and participate in social issues. |
| | Q20A5 | Through digital devices, I have ample chances to express my opinions, connect with new individuals, and expand my social network. |
| | Q20A6 | I have enhanced work and study efficiency through remote work and mobile learning. |
| | Q20A7 | I have numerous opportunities to stay in contact with acquaintances (family, relatives, friends, etc.) using digital devices. |

| AI expectation AI.expect | Q28A1 | AI technology will make my life easier. |
|---|---|---|
| | Q28A3 | Artificial intelligence technology will enhance information services. |
| | Q28A4 | I will obtain better information through the use of artificial intelligence technology. |

<Table 1> provides an overview of the items used to measure the key variables in the research model. The intention to utilize AI is assessed through four questions. The survey on AI experience targeted users who had used AI by smartphones, AI speakers, and AI voice assistants. It aimed to investigate the usefulness of these products and services in the participants' lives. One Data included four AI-equipped products and five AI services in the study. However, due to the survey being conducted in 2019, three products were selected while considering the availability of products and services at that time. Gansser & Reich (2021) presented respondents with various AI products and services to assess their experiences and perceptions, thereby measuring the acceptance of AI integrated into products.

ICT experience was assessed using seven items. Each item examined whether activities such as leisure, information retrieval, knowledge acquisition, participation, communication, and learning were performed using digital devices. Digital devices in the survey include personal computers (PCs) and mobile devices. Expectations regarding AI were measured using four indicators in the raw data. Through factor analysis in this study, it was found that the factor loading of Q28A2 was below the standard, so three items, excluding the relevant item, were utilized. Each item was rated on a 4-point scale.

<Table 2> Factor analysis

|  | Factor1 | Factor2 | Factor3 | Factor4 |
|---|---|---|---|---|
| Q20A1 | 0.431 | | | 0.199 |
| Q20A2 | 0.482 | | | 0.177 |
| Q20A3 | 0.503 | | | 0.222 |
| Q20A4 | 0.509 | | | 0.197 |
| Q20A5 | 0.551 | | | 0.239 |
| Q20A6 | 0.462 | | | 0.215 |
| Q20A7 | 0.466 | 0.125 | 0.192 | |
| Q24AC1 | | 0.329 | | |
| Q24AC2 | | 0.381 | 0.104 | |
| Q25AC3 | | 0.965 | | |
| Q28A1 | 0.212 | 0.12 | 0.447 | 0.221 |
| Q28A3 | 0.148 | | 0.558 | 0.221 |
| Q28A4 | 0.191 | | 0.671 | 0.2 |
| Q29A1 | 0.177 | 0.14 | 0.243 | 0.545 |
| Q29A2 | 0.151 | 0.113 | 0.272 | 0.562 |
| Q29A3 | 0.109 | 0.123 | 0.179 | 0.71 |
| Q29A4 | | | | 0.748 |

<Table 2> is about the outcomes of factor analysis. This analysis aims to determine whether the questionnaire items for each variable are measuring the same factor. Typically, a factor loading threshold of 0.3 or higher is used to determine the significance of the loading. The measurement items employed in this study are categorized into four factors, with factor loadings exceeding 0.3 for each item. Varimax rotation was employed as the rotation method for the factor analysis. For items belonging to the same factor, the average value was calculated and utilized as a variable.

IV. Analysis

1. Population Characteristics

The demographic characteristics of the data collected in this study are presented in <Table 3>. A total of 2,622 individuals participated, with 53.3% being male and 46.7% being female. The mean age was 34.28, and the median age was 34, indicating a relatively balanced age distribution. However, considering that the average age in Korea in 2019 was 42.2, it can be inferred that there is a higher representation of younger respondents in this study. This suggests that younger age groups have a greater exposure to AI products and services.

Occupations were organized into 6 categories. Service and sales workers is the largest group, accounting for 29.6%, while military and other occupations represented the smallest group, comprising 1.9%. In terms of educational attainment, university graduates (including graduate schools) constituted the largest segment, making up 45.4% of the sample, while elementary school graduates accounted for the lowest portion at 7.7%.

Regarding average monthly household income, 25.7% of respondents fell within the range of ₩4.00 to ₩4.99 million, followed by 23.8% falling within the range of ₩5.00 to ₩5.99 million. Considering that the average monthly household income in Korea in the fourth quarter of 2019 was approximately ₩4.77 million, it can be concluded that the sample adequately represents various income brackets without significant oversampling.

<Table 3> Demographic characteristics of data

| | |
|---|---|
| Sex | Males: 1,398 (53.3%)<br>Females: 1,224 (46.7%) |
| Age | Average age: 34.28<br>Median age: 34<br>Age variance: 188.61<br>Maximum age: 76<br>Minimum age: 7 |
| Job | Managers, professionals, office workers [job1]: 679 (25.9%)<br>Service and sales workers [job2]: 777 (29.6%)<br>Agriculture, industry, assembly, simple labor [job3]: 159 (6.1%)<br>Full-time housewives [job4]: 287 (10.9%)<br>Elementary, middle and high school students [job5]: 669 (25.5%)<br>Soldiers, others [job6]: 51 (1.9%) |

| Education | Elementary school graduation: 202 (7.7%)<br>Middle school graduation: 256 (9.7%)<br>High school graduation: 974 (37.1%)<br>College graduates: 1,190 (45.4%) |
|---|---|
| Average monthly household income | Less than ₩500,000: 3 (0.1%)<br>₩500,000-990,000: 5 (0.2%)<br>₩100-1.49 million: 12 (0.5%)<br>₩1.5-1.99 million: 23 (0.9%)<br>₩200-2.49 million: 87 (3.3%)<br>₩250-2.99 million: 151 (5.8%)<br>₩300-3.49 million: 325 (12.4%)<br>₩350-3.99 million: 337 (12.9%)<br>₩400-4.99 million: 675 (25.7%)<br>₩500-5.99 million: 625 (23.8%)<br>Over ₩6 million: 379 (14.5%) |

2. Characteristics of the number of AI uses

<Table 4> Observations by number of uses of AI

| AI product & service | no. |
|---|---|
| (1) Smartphone-mounted AI | 348(13.3%) |
| (2) AI speaker | 205(7.8%) |
| (3) AI voice assistant | 93(3.5%) |
| (1) & (2) | 198(7.6%) |
| (1) & (3) | 541(20.6%) |
| (2) & (3) | 349(13.3%) |
| (1) & (2) & (3) | 888(33.9%) |

    <Table 4> indicate how many usage of AI products and services by individual users. Among the respondents, 24.6% reported having experience with only one AI product or service, while 41.5% had used two, and 33.9% had tried all three. Among those who used multiple services, 20.6% mentioned using both a smartphone and an AI voice assistant, representing the highest proportion. In contrast, only 3.5% of respondents reported using AI voice assistants alone. The reason for this is that a large number of users primarily utilize voice assistant services such as Apple Siri, Google Assistant, and Samsung Bixby through their smartphones. Thus, it is likely that they categorized their experience as "smartphone-mounted AI" (option 1).

    33.9% of respondents had experience with all three AI products and services, suggesting a substantial number of users had explored various AI at the time. A significant majority, 75.3% of respondents, reported experiencing AI through smartphones, underscoring the prevalence of smartphone-based AI usage among users.

3. Descriptive Statistics of Key Factors

<Table 5> Descriptive Statistics of Dependent Variables, Independent Variables, and Mediator

|  | Dependent Var. | Independent Var. | | Mediator |
|---|---|---|---|---|
|  | AI.int | AI.EX | ICT.EX | AI.expect |
| Average | 2.85 | 2.97 | 3.17 | 3.13 |
| Median | 3 | 3 | 3.14 | 3 |
| Variance | 0.26 | 0.26 | 0.15 | 0.18 |
| Maximum | 4 | 4 | 4 | 4 |
| Minimum | 1 | 1 | 1 | 1 |

The average values for artificial intelligence acceptance intention (AI.int), artificial intelligence experience (AI.EX), and artificial intelligence expectation (AI.expect) are 2.85, 2.97, and 3.13, respectively. Users' acceptance intention of AI is relatively low compared to their satisfaction with AI experience and their expectations for AI. While users have experience with AI products and services and hold expectations for them, it seems that these experiences and expectations are not fully reflected in their acceptance intention. AI.int and AI.EX is smaller than the median. It suggests that the response results are more skewed towards relatively lower values. In contrast, AI.expect has a higher median value than the average, indicating that there are relatively more individuals with high expectations for artificial intelligence. The variance is the same for AI.int and AI.EX at 0.26, while AI.expect has a relatively smaller variance of 0.16. This indicates that the evaluation of AI acceptance intention and experience is distributed over a wider range compared to AI expectations. Users' acceptance intention and experience exhibit more variability compared to their expectations.

The mean of ICT experience (ICT.EX) is 3.17, and the median is 3.14. This suggests that a significant number of responses reported high ICT experience values. The sample for this study consists of individuals who have used one or more pilot AIs. Considering that it is a pilot stage, these individuals are likely to be early adopters, and it can be expected that their ICT utilization skills are relatively high compared to other groups.

[Figure 2] Scatter diagram of AI experiences and AI expectations

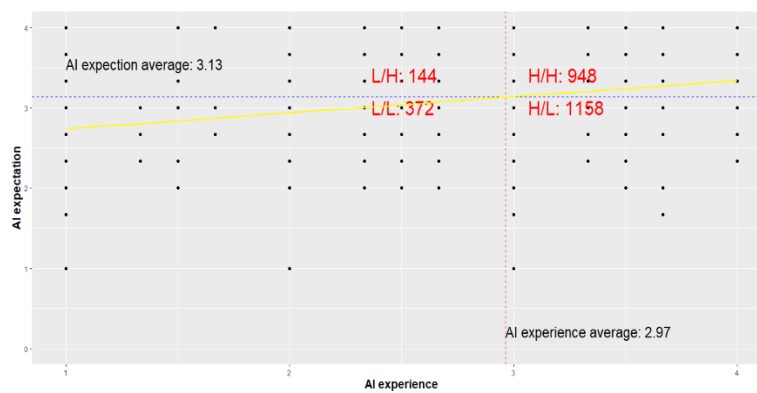

[Figure 3] Scatter diagram of ICT experiences and AI expectations

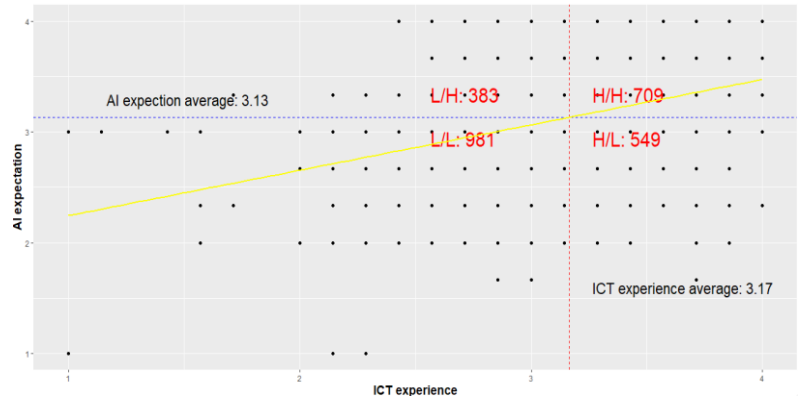

[Figure 4] Scatter diagram of artificial intelligence experience and acceptance intention

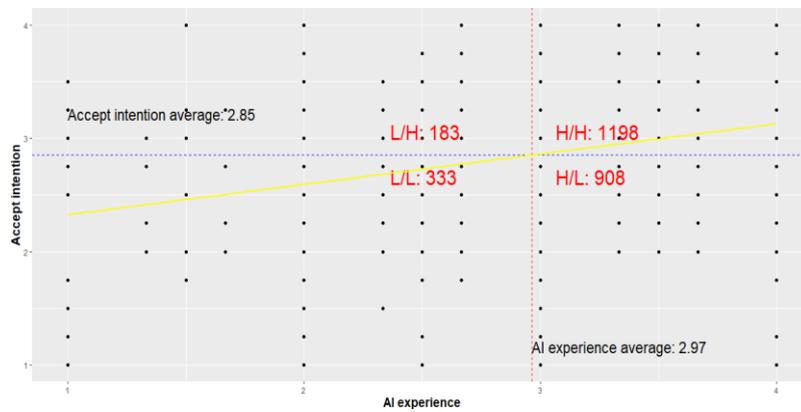

[Figure 4] Scatter diagram of artificial intelligence experience and acceptance intention

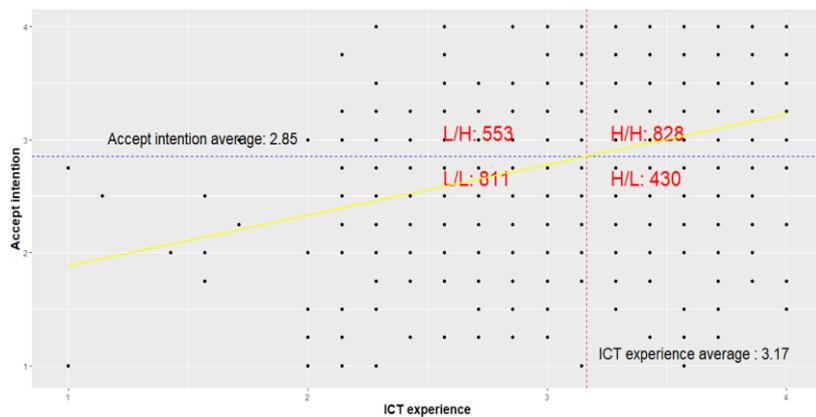

Figures 2, 3, 4, and 5 illustrate the distribution of each factor and the relationships between them. In [Figure 2], a scatter plot displays the relationship between AI.EX and AI.expect. The distribution of AI.EX spans a relatively wide range compared to ICT.EX. Around 80.3% of respondents have AI.EX scores higher than the average. Among them, a larger proportion falls into a group where AI.expect is relatively lower compared to AI.EX. A majority of respondents are realistically adjusting their future expectations through AI experiences.

[Figure 3] depicts a scatter plot showing the relationship between ICT.EX and AI.expect. The

data indicates that satisfaction with ICT experience is more clustered compared to satisfaction with AI experience. Notably, the group with low ICT experience and low AI expectations comprises the highest proportion (37.4%) among the four groups examined. This finding can be attributed to the fact that most artificial intelligence products and services are offered in conjunction with existing ICT. Additionally, information about artificial intelligence is often sought through digital media. Consequently, individuals with limited ICT utilization tend to have relatively lower expectations for artificial intelligence.

[Figure 4] illustrates the relationship between AI.EX and AI.int. When individuals reported a highly satisfactory experience with artificial intelligence, 45.7% of them expressed a high intention to accept artificial intelligence, while 34.6% did not. It is important to note that the experience users gain from AI products and services in the pilot stage is limited. When making decisions about AI acceptance, various factors come into play, including cost, living environment, risk, and interpersonal relationships. Therefore, even with high satisfaction, individuals may still have varying levels of willingness to accept AI based on these additional factors.

A notable characteristic in [Figure 5] is that AI.int is likely to be low in the group with low ICT.EX, while it is likely to be high in the group with high ICT.EX. This suggests that individuals who do not utilize existing ICT in their daily lives or have faced difficulties in doing so may perceive AI as unhelpful or challenging to use, leading to a lower intention to accept it. Conversely, individuals who have experience using existing ICT to enhance efficiency and convenience in their lives and work are more likely to perceive AI as beneficial. Consequently, they tend to form a higher intention to accept artificial intelligence. This finding suggests that prior experience with ICT plays a significant role in shaping individuals' attitudes and intentions towards accepting AI.

4. Regression analysis results

The analysis regression analysis results of the study are presented in <Table 6>. The study employed three regression models to investigate the mediating effect. The primary variables in the model are the independent variables, AI.EX and ICT.EX, while the mediator is AI.expect. The dependent variable is AI.int.

Model 1 focuses on examining the influence of the independent variables (AI.EX, ICT.EX) on the dependent variable (AI.int). Model 2 explores the effect of the independent variables on the dependent variable (AI.int), excluding the mediator (AI.expect). Model 3 includes both the independent variables and the mediator, analyzing their effects on the dependent variable.

The goodness of fit of the three models can be assessed through the F value. In Model 1, the F value is 50.82 ($p < 0.05$), in Model 2, the F value is 49.57 ($p < 0.05$), and in Model 3, the F value is 79.44 ($p < 0.05$). Generally, when the p-value associated with the F value is less than 0.05, it indicates that the model has a significant goodness of fit.

The explanatory power of each model can be evaluated using the adjusted R-squared. Model 1 explains 18.6% of the variance in AI.expect (Adjusted R-squared: 0.186). Model 2 explains 18.2% of the variance in AI.int (Adjusted R-squared: 0.182). Model 3, which includes both the independent variables and the mediator, explains 28% of the variance in AI.int (Adjusted R-squared: 0.28). The improvement in the explanatory power of Model 3 compared to Model 2 can be attributed to the inclusion of artificial intelligence expectation (AI.expect) in the model. This indicates that expectations for AI play a crucial role in explaining changes in the intention to accept AI.

<Table 6> Regression analysis results

| Model | Model 1 | | | Model 2 | | | Model 3 | | |
|---|---|---|---|---|---|---|---|---|---|
| DV | AI.expect | | | AI.int | | | AI.int | | |
| statistics | β | SE | t | β | SE | t | β | SE | t |
| (Intercept) | 1.62 *** | 0.1 | 16.58 | 1.13 *** | 0.12 | 9.56 | 0.45 *** | 0.12 | 3.87 |
| AI.EX | 0.15 *** | 0.01 | 9.96 | 0.21 *** | 0.02 | 11.75 | 0.15 *** | 0.02 | 8.67 |
| ICT.EX | 0.35 *** | 0.02 | 17.08 | 0.36 *** | 0.02 | 14.59 | 0.21 *** | 0.02 | 8.76 |
| AI.expect | | | | | | | 0.42 *** | 0.02 | 18.89 |
| num.use | 0.03 *** | 0.01 | 2.88 | 0.01 | 0.01 | 0.61 | 0 | 0.01 | -0.42 |
| age | -0.00 *** | 0 | -4.03 | -0.00 *** | 0 | -4.67 | -0.00 *** | 0 | -3.48 |
| female | -0.01 | 0.02 | -0.64 | -0.08 *** | 0.02 | -4.01 | -0.08 *** | 0.02 | -4.04 |
| job[2] | -0.01 | 0.02 | -0.6 | -0.06 ** | 0.03 | -2.56 | -0.06 ** | 0.02 | -2.51 |
| job[3] | -0.08 ** | 0.04 | -2.31 | -0.11 *** | 0.04 | -2.59 | -0.08 * | 0.04 | -1.9 |
| job[4] | -0.01 | 0.03 | -0.42 | 0.03 | 0.04 | 0.7 | 0.03 | 0.03 | 0.9 |
| job[5] | -0.06 * | 0.04 | -1.69 | -0.06 | 0.04 | -1.34 | -0.03 | 0.04 | -0.8 |
| job[6] | 0 | 0.06 | 0.02 | 0.02 | 0.07 | 0.32 | 0.02 | 0.06 | 0.33 |
| education | -0.01 | 0.01 | -0.48 | 0.01 | 0.02 | 0.79 | 0.02 | 0.01 | 1.02 |
| income | 0.01 ** | 0 | 2.12 | 0.02 *** | 0.01 | 3.24 | 0.01 *** | 0.01 | 2.67 |
| Obs | 2622 | | | 2622 | | | 2622 | | |
| F | 50.82 on 12 and 2609 DF | | | 49.57 on 12 and 2609 DF | | | 79.44 on 13 and 2608 DF | | |
| R2/ R2 adj | 0.189 / 0.186 | | | 0.186 / 0.182 | | | 0.284 / 0.280 | | |
| * p<0.1 / ** p<0.05 / *** p<0.01 | | | | | | | | | |

Based on the analysis results, the hypotheses presented in this study were tested as follows:

Hypothesis 1 (H1): The higher the evaluation of AI product and service experience (AI.EX), the better the AI acceptance intention (AI.int). This hypothesis is supported by the results in Model 2 (β=0.21, p<0.01) and Model 3 (β=0.15, p<0.01). The coefficient (β) of AI.EX decreases in Model 3 because some effects are mediated by AI expectations (AI.expect) and impact the acceptance intention.

Hypothesis 2 (H2): The higher the evaluation of ICT experience (ICT.EX), the better the AI acceptance intention (AI.int). This hypothesis is supported by the results in Model 2 (β=0.36, p<0.01) and Model 3 (β=0.21, p<0.01). Similar to the relationship between AI.EX and AI.int, some effects are mediated by AI expectations and influence acceptance intention. Therefore, the coefficient (β) of ICT.EX in Model 3 is lower than that in Model 2.

Hypothesis 3 (H3): The higher the expectation for AI (AI.expect), the better the intention to accept AI (AI.int). Model 3 suggests a significant positive relationship (β=0.42, p<0.01) between AI.expect and AI.int, supporting this hypothesis.

In summary, the analysis results confirm the hypotheses. The evaluation of product and service experience (AI.EX) and ICT experience (ICT.EX) have positive effects on AI acceptance intention (AI.int), and these effects are partly mediated by AI expectations (AI.expect). Furthermore, higher expectations for AI (AI.expect) are associated with a better intention to accept AI (AI.int).

Hypothesis 4 (H4): The AI (AI.expect) partially mediate the relationship between the evaluation of AI experience (AI.EX) and ICT experience evaluation (ICT.EX) and AI acceptance intention (AI.int). Baron & Kenny (1986) introduced a method to confirm the mediating effect, which involves several conditions. First, the independent variable should have a significant effect on the mediator. Second, the independent variable should have a significant effect on the dependent variable. Third, the mediator should have a significant effect on the dependent variable. Last, the effect of the independent variable on the dependent variable in the model with the mediator should be smaller compared to the effect of the independent variable on the dependent variable in the model without the mediator. By examining these conditions, the study aims to establish the mediating effect of expectations of artificial intelligence on the relationship between AI.EX, ICT.EX, and AI.int.

Among the proposed models, Model 1 examined the effect of the independent variables on the mediator. Both AI.EX and ICT.EX showed significant effects on the mediator AI.expect (AI.EX: 1.5, p<0.01; ICT.EX: 0.35, p<0.01). In Model 2, which excluded the mediator, AI.EX and ICT.EX had a significant effect on the dependent variable AI.int (AI.EX: 0.21, p<0.01; ICT.EX: 0.36, p<0.01). Model 3, which included the mediator, demonstrated that AI.EX, ICT.EX, and AI.expect all had significant effects on AI.int (AI.EX: 0.15, p<0.01; ICT.EX: 0.21, p<0.01; AI.expect: 0.42, p<0.01). Notably, when comparing the coefficients of AI.EX and ICT.EX between Model 2 and Model 3, they were found to have decreased (AI.EX: 0.21 -> 0.15; ICT.EX: 0.36 -> 0.21) in Model 3.

To test the statistical significance of the mediating effect, the Sobel test, Aroian test, and Goodman test were conducted. Since this study assumed two mediating effect pathways, one from AI.EX to AI.expect and the other from ICT.EX to AI.expect, the three tests were conducted separately for each pathway. The null hypothesis for all three tests was that there is no mediating effect. The results of the three tests for both pathways confirmed that the mediating effect was statistically significant. The results are presented in <Table 7>. Additionally, [Figure 6] provides a schematic representation of the relationships between the main factors in the study.

<Table 7> Mediating effect statistical test results

| Path | Test | Test statistic | Std. Error | p-value |
|---|---|---|---|---|
| AI.EX -> AI.expect -> AI.int | Sobel test | 8.81 | 0.007 | 0 |
| | Aroian test | 8.8 | 0.007 | 0 |
| | Goodman test | 8.82 | 0.007 | 0 |
| ICT.EX -> AI.expect -> | Sobel test | 7.8 | 0.009 | 0 |
| AI.int | Aroian test | 7.79 | 0.009 | 0 |
| | Goodman test | 7.81 | 0.009 | 0 |

[Figure 6] Relationship between major factors

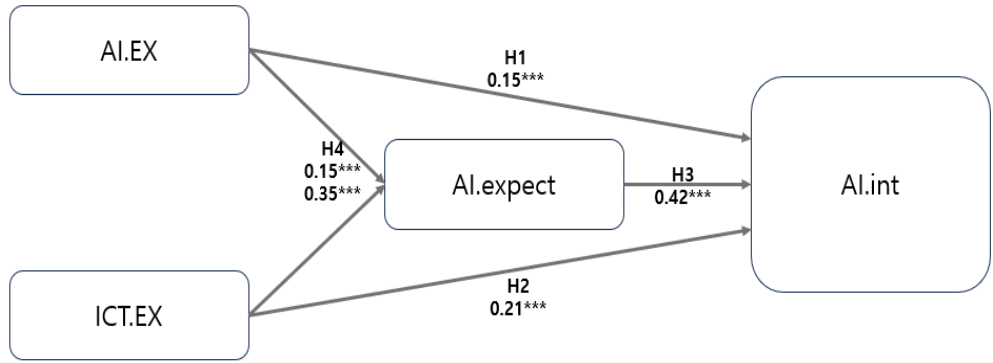

## 5. Discussion

In this study, we analyzed the impact of AI usage experience, ICT usage experience, and expectations of AI on the intention to accept AI. The findings revealed that a higher intention to accept AI was associated with the perception that the available AI products and services were beneficial to daily life. Through experiences with AI, users were able to gain detailed knowledge about the technology's characteristics and applications. They were exposed to information about the vision and future development of artificial intelligence, and had opportunities to compare the performance and cost of different AI offerings.

There exists a positive relationship between the evaluation of AI experience and the intention to accept it. Knowledge acquired from past experiences tends to be more memorable, and experienced events have a greater influence on the formation of intentions. Poor current experiences with AI products and services can have a negative impact on future acceptance.

This observation aligns with the general acceptance process of various digital products and services, where positive service experiences tend to promote higher intention to accept and adopt new technologies. For example, individuals who have experienced using inconvenient and inefficient kiosk services in the past may have lower acceptance intentions, even if usability improves over time. On the other hand, users who have had convenient experiences with wireless earphones are more likely to have a positive acceptance intention toward other wireless products, and they are more inclined to actively adopt such technologies.

According to the findings of this study, not only AI products and services, but also experiences with other ICT technologies influence the intention to accept AI. Individuals who have had satisfactory experiences with ICT across various activities tend to have a higher intention to accept products and services equipped with AI. Additionally, information about AI, related products and services, and their evaluations is predominantly disseminated through the Internet. Therefore, individuals who actively use ICT are more likely to be exposed to such information, which can further increase their intention to accept AI. Previous research in the field of education suggests that former ICT usage experience can serve as a source of confidence in using AI. However, if individuals have low current levels of ICT usage or repeatedly encounter unsatisfactory experiences, it can lead to a lower intention to accept AI.

Expectations for AI play a direct role in shaping the intention to accept it. When the variable measuring expectations is excluded from the model, the explanatory power of the model significantly diminishes. In this study, it was found that the more positive the expectations for AI, the higher the acceptance intention. AI experiences have a direct impact on the intention to accept, while also influencing it through expectations. Approximately 30% of the effects of AI experiences influence acceptance intention through expectations. Positive experiences with artificial intelligence enhance expectations and foster a greater intention to accept it.

ICT experiences also directly affect the intention to accept AI, with some effects being mediated by expectations. Approximately 42% of the effects of ICT experiences were found to be mediated by AI expectations. Past experiences with ICT products and services play a significant role in shaping expectations for AI and promoting the willingness to accept it.

Expectations of AI can be understood as a set of beliefs and perceptions held about AI technology. Conceptually, expectations are formed prior to the actual usage of a technology, product, or service, but they can be adjusted based on direct and indirect experiences. In this study, both AI experiences ($\beta = 0.15$, $p < 0.01$) and ICT experiences ($\beta = 0.15$, $p < 0.01$) were found to have a significant effect on expectations. These findings emphasize the importance of managing current experiences with AI products and services, as well as ICT experiences, to foster positive expectations for artificial intelligence.

V. Conclusion

As AI technology becomes increasingly accepted in society, various efforts will be made to apply it across different fields. However, if there is a low intention to accept AI among users, it may result in low adoption of AI. This could lead to a decrease in demand for AI products and services, an increase in distrust, and a slowdown in technological innovation. Previous theories and studies on AI acceptance have focused on identifying acceptance intentions and exploring the factors that influence intention. However, there are several criticisms of these existing theories and studies.

In previous theories and studies, experience with technology is often identified as a moderating variable rather than having a direct impact on the acceptance intention. However, empirically, experience plays a crucial role in shaping knowledge and perceptions of technology, significantly. While there are active discussions about the use of AI, its application in various fields, and its impact on daily life, the precise scope and influence of AI remain uncertain. In this context, the experience of using artificial intelligence products and services becomes a crucial foundation for evaluating the usefulness and ease of use of AI. Therefore, this study examines how current experiences with artificial intelligence influence the intention to accept artificial intelligence in the future. Based on the analysis results of this study, it is found that higher evaluations of AI products and services in terms of their helpfulness in daily life, as well as greater satisfaction with ICT experiences, are associated with higher expectations for AI. Expectations for AI, in conjunction with experiences with AI and ICT, play a crucial role in shaping the intention to accept AI.

Based on the findings of this study, the following suggestions can be offered to companies and public organizations planning to introduce artificial intelligence in the future:

Manage the user experience of existing digital services: It is crucial to ensure that users have a positive experience with the currently provided digital services. Users who have satisfactory experiences with existing digital services tend to have higher expectations and acceptance intentions for artificial intelligence. When users perceive existing digital services as efficient, convenient, and helpful, they expect that the introduction of artificial intelligence will further enhance these services, leading to increased acceptance. Moreover, positive experiences with digital services can help alleviate fears and resistance towards artificial intelligence.

Manage the user experience of pilot artificial intelligence products and services: Care must be taken when providing artificial intelligence-enabled products and services on a pilot basis. Indiscriminately deploying pilot AI services and subjecting users to repeated unsatisfactory experiences can diminish expectations for AI, intensify skepticism about its necessity, and undermine acceptance. Even in the event of pilot service failures, it is important to communicate the potential for development and improvement to users.

Provide users with detailed information about the performance, convenience, and benefits of utilizing AI: Provide users with realistic and specific information, avoiding vague promises or exaggerated claims. Unrealistic expectations can lead to disappointment if the actual experiences with artificial intelligence fail to meet these inflated expectations, which may result in decreased acceptance of AI.

By implementing these suggestions, companies and public organizations can better manage user experiences, address expectations, and foster positive attitudes towards AI, thus facilitating its successful integration into various domains.

This study has several limitations that should be noted. First, it focused on analyzing acceptance intention rather than actual acceptance of AI. While acceptance intention is an important factor in technology acceptance, it is not the sole determinant. To comprehensively understand the factors influencing actual acceptance of artificial intelligence, other factors such as cost, support, policies, legal systems, and social context need to be considered. The impact of artificial intelligence experience, ICT experience, and expectations may operate differently when analyzing actual acceptance of artificial intelligence. Second, this study did not encompass a wide range of AI experiences. The data used in this study was collected in 2019 and focused on artificial intelligence experiences related to commonly available AI products and services at that time. With the recent emergence of generative artificial intelligence and the increasing diversity of AI products and services, it is important for future studies to develop indicators that cover a broader range of AI experiences. Third, the relationship between AI experience and ICT experience was not explicitly examined in this study. There may exist a pathway where ICT experience promotes AI experience, which in turn influences acceptance intention. Exploring this relationship could provide further insights into the interplay between AI and ICT experiences.

VI. Acknowledgments

This work was supported by the Ministry of Education of the Republic of Korea and the National Research Foundation of Korea (NRF-2021S1A5C2A03087287).